\documentclass[aps,prl,reprint,amsmath,amssymb,showpacs,superscriptaddress]{revtex4-1}
\usepackage{CJK}
\usepackage{graphicx}
\usepackage{dcolumn}
\usepackage{bm}

\usepackage{color}
\usepackage{hyperref}

\renewcommand{\vec}[1]{\mbox{\boldmath $#1$}}

\begin{document}


\title{Non-local orbital-free density functional theory incorporating nuclear shell effects}

\author{Xinhui Wu}
\email{wuxinhui@fzu.edu.cn}
\affiliation{Department of Physics, Fuzhou University, Fuzhou 350108, Fujian, China}

\author{Gianluca Col\`o}
\email{Gianluca.Colo@mi.infn.it}
\affiliation{Dipartimento di Fisica, Universit\`a
degli Studi di Milano, via Celoria 16, 20133 Milano, Italy,}
\affiliation{INFN, Sezione di Milano, via Celoria 16, 20133 Milano, Italy}

\author{Kouichi Hagino}
\email{hagino.kouichi.5m@kyoto-u.ac.jp}
\affiliation{ 
Department of Physics, Kyoto University, Kyoto 606-8502,  Japan}
\affiliation{Institute for Liberal Arts and Sciences, Kyoto University, Kyoto 606-8501, Japan}
\affiliation{ 
RIKEN Nishina Center for Accelerator-based Science, RIKEN, Wako 351-0198, Japan
}

\author{Pengwei Zhao}
\email{pwzhao@pku.edu.cn}
\affiliation{State Key Laboratory of Nuclear Physics and Technology, School of Physics, Peking University, Beijing 100871, China}


\begin{abstract}
Incorporating nuclear shell effects within the framework of orbital-free density functional theory (DFT) has remained a longstanding challenge in nuclear physics. 
While the Hohenberg-Kohn theorem formally guarantees the existence of an orbital-free density functional 
that is capable of describing all many-body effects, including shell effects, practical attempts since the 1970s have consistently failed to capture such effects. 
This persistent difficulty has even led to the misconception that the orbital-free DFT is inherently unable to describe nuclear shell effects. 
Here we develop a {\it non-local} orbital-free DFT approach for atomic nuclei and demonstrate that nuclear shell effects can be successfully incorporated into the orbital-free DFT through the construction of a non-local 
kinetic energy density functional. 
In particular, we show that the non-local orbital-free functional yields a nucleon localization function that, as an established indicator of shell effects, exhibits consistent behavior with the exact Kohn-Sham solution. 

\end{abstract}


\maketitle



{\it Introduction.}
Research on quantum many-body systems is essential across a wide range of scientific fields. 
Directly solving the quantum many-body Schr{\"o}dinger equation exhibits exponential computational complexity with increasing number of particles, and there are still long-standing challenges to obtain the solutions for 
systems with large numbers of particles. 
The density functional theory (DFT), based on the Hohenberg-Kohn theorem~\cite{Hohenberg1964Phys.Rev.},  provides fully quantum solutions at a fraction of the cost of directly solving the Schr{\"o}dinger equation by mapping the coupled many-body problem to a single-particle problem.
The DFT has been extensively investigated
in various fields, e.g., quantum chemistry, condensed-matter physics, and nuclear physics\cite{jones2015}. 

In nuclear physics, the DFT has been widely applied~\cite{Ring1996PPNP, Bender2003Rev.Mod.Phys., Vretenar2005Phys.Rep.101, Erler2012Nature, Nakatsukasa2016Rev.Mod.Phys., Meng2016book, Shen2019PPNP, schunck2019energy, Robledo2019,Colo2020nuclear} predominantly within the Kohn-Sham (KS) scheme~\cite{Kohn1965Phys.Rev.}, which introduces auxiliary one-body orbitals to compute the kinetic energy.
In contrast, in the scheme of orbital-free DFT (OF-DFT)~\cite{Levy1984PRA}, one aims to express the energy solely as a functional of the density.
The OF-DFT returns to the roots of the Hohenberg-Kohn theorem and puts the density back into centrality in the DFT study. 
This makes the OF-DFT a more fundamental framework than the Kohn-Sham DFT, facilitating a deeper understanding of the Hohenberg-Kohn theorem~\cite{Mi2023CR}.
The solution of OF-DFT is in principle very simple and quick to obtain, as there is only one ``orbital", i.e., the square root of the density distribution.
Therefore, the approach is very attractive for cases where computations become demanding in the Kohn-Sham scheme, such as superheavy nuclei, or nuclei coexisting with a neutron ``sea'' in the inner crust of neutron stars. 

The motivations from both theoretical and practical aspects have made the OF-DFT a longstanding ideal pursuit for many DFT researchers.
The stumbling block of the OF-DFT is how to seek sufficiently accurate descriptions of the kinetic energy with the density alone, since the Hohenberg-Kohn theorem only proves the existence of energy density functional but does not provide its actual form.
The exact kinetic energy density functional is currently known for only two cases.
One case is for a one-particle system, which is the von Weizs{\"a}cker (vW) kinetic energy density functional (KEDF)~\cite{Weizsaecker1935Z.Physik}. 
Note that the vW kinetic functional can also be regarded as an exact KEDF for the ground states of 
many-{\it boson} systems.
The other case is for a uniform noninteracting system, which is the Thomas-Fermi (TF) KEDF~\cite{Thomas1927, Fermi1927} derived from the local implementation of a uniform Fermi-gas model. 
These two KEDFs as well as their extensions and combinations have been widely used in practical nuclear structure calculations~\cite{Bethe1968PR, Brack1972Rev.Mod.Phys., Bohigas1976Phys.Lett.B, Brack1985Phys.Rep., Dutta1986Nucl.Phys.A, Myers1990AP, Aboussir1995Atom.DataNucl.DataTables, Centelles1990Nucl.Phys.A, Centelles2007Ann.Phys., Bulgac2018Phys.Rev.C, Colo2023PTEP}.
However, in these previous attempts with the orbital-free approach, 
all nuclei were spherical in their ground states~\cite{Brack1985Phys.Rep., Colo2023PTEP} and the obtained ground-state densities were smooth, due to the lack of quantum shell effects~\cite{Brack1985Phys.Rep., Centelles1990Nucl.Phys.A, Centelles2007Ann.Phys.}.

From a microscopic perspective, the quantum shell effects are related to a non-uniform distribution of single-particle energies and, thus, an accurate description of the quantum shell effects by the orbital-free DFT has remained an elusive topic for nuclear physics. 
In principle, it is guaranteed by the Hohenberg-Kohn theorem that the quantum shell effects can be described by the OF-DFT approach.
However, in practice, this has not yet been achieved despite many attempts since 1970s, in which one always needed to supplement additional shell corrections~\cite{Strutinsky1967NPA, Strutinsky1968NPA, Brack1972Rev.Mod.Phys., Brack1973Nucl.Phys.A, Bohigas1976Phys.Lett.B, Chu1977Phys.Lett.B} to incorporate the shell effects. 
A feasible route to incorporating nuclear shell effects into a pure OF-DFT has remained unclear for a long time.
As a result, this even gives rise to a common misconception that orbital-free functionals cannot describe the nuclear shell effect.

It is worth noting that this challenge is not unique to nuclear systems~\cite{Finzel2015}. 
For electronic systems, the accurate incorporation of shell-like effects, such as atomic shell structure and shell filling in clusters, within the OF-DFT has also been proven difficult~\cite{Hodges1973CJP, Tal1978IJQC, Garca1996PhysRevA.54.1897, Thompson2020PhysRevA.102.012813, Mi2023CR}.
Quite a few efforts in this direction have been made~\cite{Chacon1985PhysRevB.32.7868, Wang1992PRB, Garca1996PhysRevB.53.9509, Garca1998PhysRevB.57.4857, Wang1999PhysRevB.60.16350,Mi2023CR,witt2018}, which can serve as references for nuclear physics research.

Recent advances in constructing kinetic energy density functionals using machine learning techniques have practically demonstrated promising success in capturing nuclear shell effects within the OF-DFT framework~\cite{Wu2022Phys.Rev.C, Hizawa2023PRC, Chen2024IJMPE, Wu2025CP}, rekindling a 
hope for developing OF-DFT with quantum shell effects.
While the machine-learning-based OF-DFT has shown practical success in capturing the nuclear shell effects, the black-box nature of these complex, high-dimensional functionals limits their physical interpretability.
This has left a critical gap in understanding how to incorporate quantum shell effects into 
the OF-DFT between the practical and theoretical perspectives.

The aim of this Letter is to bridge this critical gap. 
Our key idea is to introduce a non-locality to the kinetic energy functional and to use the linear response theory to determine it \cite{Wang1992PRB}. 
We shall demonstrate that this approach captures well the shell effects, offering a transparent and interpretable orbital-free method to understand nuclear quantum shell effects.


{\it Formalism of non-local OF-DFT. }
The aim of OF-DFT is to establish the kinetic energy solely as a functional of the local density, $\rho(\vec{r})$, 
that is,
\begin{equation}
    T[\rho]=\frac{\hbar^2}{2m}\int d\vec{r}\ \tau[\rho], 
\end{equation}
where $m$ is the nucleon mass and $\tau[\rho]$ is the kinetic energy density.
The OF-DFT can exist in local, semi-local, and non-local form.
Local OF-DFT means that the kinetic energy density $\tau({\bm r})$ depends only on the density $\rho(\bm{r})$ 
at the same point like, e.g., in the case of the TF functional given by
\begin{equation}\label{Eq:TTF}
  T^{\rm TF}[\rho]=\frac{\hbar^2}{2m}\int {\rm d}^3 r \frac{3}{5}(3\pi^2)^{2/3} \rho^{5/3}({\bm r}).  
\end{equation}
The semi-local OF-DFT adds spatial derivatives of the density like in the case of the extended TF (ETF) functionals, or when the TF functional is combined with the vW functional in the widely used form
\begin{equation}\label{Eq:TTFvW}
    T^{\rm TFvW}[\rho]=\alpha  T^{\rm TF}[\rho] + \beta T^{\rm vW}[\rho],
\end{equation}
where $\alpha$ and $\beta$ are constants and the vW functional $T^{\rm vW}[\rho]$ is given by 
\begin{equation}
T^{\rm vW}[\rho]= \frac{\hbar^2}{2m}\int {\rm d}^3 r\, (\nabla\sqrt{\rho})^2.
\end{equation}
The non-local OF-DFT most generally has the form,
\begin{align}\label{Eq:mbodykf}
  T^{\rm nl}[\rho]=&\int {\rm d}^3r_1 {\rm d}^3r_2...{\rm d}^3r_m\, \rho^{\alpha_1}({\bm r}_1)\rho^{\alpha_2}({\bm r}_2) \notag \\
  &\times ...\rho^{\alpha_m}({\bm r}_m) f({\bm r}_1,{\bm r}_2,...,{\bm r}_m),  
\end{align}
where $f({\bm r}_1,{\bm r}_2,...,{\bm r}_m)$ is an $m$-point kernel to be determined. 
For a slowly varying density, one can expect that the kinetic energy returns to the TF form~\eqref{Eq:TTF}, 
which leads to a condition of 
$\sum_i\alpha_i = 5/3$ if $f$ is independent of $\rho$.
The form~\eqref{Eq:mbodykf} can be traced back to the original paper by Hohenberg-Kohn~\cite{Hohenberg1964Phys.Rev.} as a formal expansion of the exact density functional.
So far, to the best of our knowledge, the form \eqref{Eq:mbodykf} has not been used for $m>3$ and only once for $m=3$~\cite{Wang1992PRB}.

Here, we follow the derivation in Ref.~\cite{Wang1992PRB}, and focus on the two-point $(m=2)$ functional case, that is,
\begin{equation}\label{Eq:Tnl}
  T^{\rm nl}[\rho]=\frac{\hbar^2}{2m}\int {\rm d}^3 r \int {\rm d}^3 r' \rho^{5/6}({\bm r})f({\bm r},{\bm r}') \rho^{5/6}({\bm r}'). 
\end{equation}
Here, the two-point kernel $f({\bm r},{\bm r}')$ is written as $f(k_F|{\bm r}-{\bm r}'|)$, where $k_F$ is the Fermi momentum for a typical density $\rho_0$ of the system, and it is introduced to render the argument of the kernel dimensionless.
One may take $k_F$ as a constant $k_F=(3\pi^2\rho_0)^{1/3}$, and adjust $\rho_0$ as a free parameter. We first notice that the non-local functional~\eqref{Eq:Tnl} should reduce to the TF functional~\eqref{Eq:TTF} when the typical length of variation of 
the density $\rho({\bm r})$ is larger than the range of $f(k_F|{\bm r}-{\bm r}'|)$. This leads to the condition
\begin{equation}\label{Eq:TFlim}
  \int {\rm d}^3 r' f(k_F|{\bm r}-{\bm r}'|) = \frac{3}{5}(3\pi^2)^{2/3}.
\end{equation}
A second condition to determine the form of $f(k_F|{\bm r}-{\bm r}'|)$ can then be obtained from the linear-response theory, 
which relates a perturbation in the effective potential $\delta V({\bm r})$ to the
corresponding change of the density $\delta\rho({\bm r})$.
To this end, let us express 
the total energy energy of the system $E_{\rm tot}$ as 
\begin{equation}\label{Eq:Etot}
  E_{\rm tot}[\rho] = T^{\rm nl}[\rho] + E_{\rm int}[\rho] 
\end{equation}
where $E_{\rm int}[\rho]$ is the interaction term. 
With the condition that  
\begin{equation}
E'_{\rm tot}[\rho] = E_{\rm tot}[\rho] - \mu \int {\rm d}^3\ r \rho,
\end{equation}
where 
$\mu$ is the Lagrange multiplier ensuring the conservation of the total particle number,
is stable with respect to a small variation of $\rho$,
one obtains
\begin{equation}\label{Eq:dEtot}
  \frac{\hbar^2}{2m} \frac{5}{3}\rho^{-1/6}({\bm r})\int {\rm d}^3 r' f(k_F|{\bm r}-{\bm r}'|) \rho^{5/6}({\bm r}') + \delta V -\mu = 0, 
\end{equation}
where $\delta V$ is defined as $\delta V\equiv \frac{\delta E_{\rm int}}{\delta\rho}$.

Assuming that the auxiliary $\delta V$ is weak, so that the linear-response limit can be employed and the density variation $\delta\rho({\bm r})=\rho({\bm r})-\rho_0$ is small, 
one obtains 
\begin{align}\label{Eq:lrrhov}
  0=&\frac{\hbar^2}{2m} \frac{5}{3}\rho^{-1/6}_{0}\cdot\left[1-\frac{1}{6}\frac{\delta\rho({\bm r})}{\rho_{0}}\right] \notag \\
  &\times \int {\rm d}^3 r' f(k_F|{\bm r}-{\bm r}'|) \rho^{5/6}_{0}\cdot\left[1+\frac{5}{6}\frac{\delta\rho({\bm r}')}{\rho_{0}}\right] + \delta V -\mu, 
\end{align}
to linear order in $\delta V$ and $\delta\rho$.
The zero-order terms of this equation lead to Eq.~\eqref{Eq:TFlim} again.
The first-order terms give
\begin{align}\label{Eq:lrrfod}
  &\frac{\hbar^2}{2m}\frac{25}{18} \rho_0^{-1/3} \int {\rm d}^3 r' f(k_F|{\bm r}-{\bm r}'|) \delta\rho({\bm r}') \notag \\
  &- \frac{\hbar^2}{2m}\frac{5}{18}\rho_0^{-1/3}F(0)\delta\rho({\bm r}) = -\delta V,
\end{align}
where $F(0)\equiv\int {\rm d}^3 r' f(k_F|{\bm r}-{\bm r}'|) = \frac{3}{5}(3\pi^2)^{2/3}$ [see Eq. (\ref{Eq:TFlim})].
Transforming this equation to the momentum space,
one obtains
\begin{equation}\label{Eq:lrrfodq}
  \delta V(\bm q) = \frac{5}{18}\frac{\hbar^2}{2m}\rho_0^{-1/3}F(0) \delta\rho({\bm q}) - \frac{25}{18} \frac{\hbar^2}{2m} \rho_0^{-1/3} F({\bm q})\delta\rho({\bm q}).
\end{equation}
In the linear response theory, $\delta\rho$ and $\delta V$ are related to each other with 
the response function $G({\bm q})$ as
\begin{equation}\label{Eq:lrsq}
  \delta\rho({\bm q}) = -G({\bm q})\delta V({\bm q}). 
\end{equation}
For unpolarized, uniform and non-interacting Fermi systems, 
the response function $G({\bm q})$
is given by the Lindhard function~\cite{Lindhard1954lf}, which in its static limit reads 
\begin{align}
  &-G({\bm q}) = -\frac{mk_{F}}{\pi^2\hbar^2}l(x), \label{Eq:lind1} \\
  &l(x) = \frac{1}{2}\left( 1+\frac{1-x^2}{2x}\log\left|\frac{1+x}{1-x}\right| \right),  \label{Eq:lind2}
\end{align}
with $x = q/2k_{F}$.
Combining
Eqs.~\eqref{Eq:lrrfodq}, \eqref{Eq:lrsq}, \eqref{Eq:lind1} and \eqref{Eq:lind2}, one obtains
\begin{equation}\label{Eq:kerq}
   F({\bm q}) = \frac{12}{25} (3\pi^2)^{2/3} \left[  \frac{1}{l(q/2k_F)} + \frac{1}{4} \right].
\end{equation}

One can obtain the original kernel in the coordinate space, $f(k_F|{\bm r}-{\bm r}'|)$, by taking 
the inverse Fourier transform of $F({\bm q})$. To this end, it is convenient to define 
\begin{equation}
  l_1(x)=\frac{5}{8}\left[\frac{1}{l(x)} - 3x^2 + \frac{3}{5}\right], 
\end{equation}
which goes to zero as $x$ goes to infinity, that is, $l_1(x) \to 0$ as $x \to \infty$.
The coefficient 5/8 guarantees 
the normalization condition $\int d\vec{r}\,f_1(\vec{r})=1$, where $f_1(\vec{r})$ is the inverse Fourier transform of $l_1(x)$. 
With $l_1(x)$, $F({\bm q})$ in Eq. (\ref{Eq:kerq}) is expressed as 

\begin{equation}
   F({\bm q}) = \frac{12}{25}(3\pi^2)^{2/3}  \left[ \frac{8}{5}l_1(q/2k_F) + 3\frac{q^2}{4k_F^2} - \frac{7}{20} \right], 
\end{equation}
from which one obtains  
\begin{equation}
   f(k_F|{\bm r}-{\bm r}'|) = \frac{12}{25} (3\pi^2)^{2/3} \left[ \frac{8}{5}f_1(k_FR) - \frac{3}{4k_F^2}\nabla^2 - \frac{7}{20}\delta(R) \right],\label{Eq:kerinr}
\end{equation}
where $R\equiv |{\bm r}-{\bm r}'|$.
Inserting Eq.~\eqref{Eq:kerinr} into Eq.~\eqref{Eq:Tnl}, one can finally have
\begin{equation}\label{Eq:Tnlfinal}
\begin{aligned}
  T^{\rm nl}[\rho]=&\frac{\hbar^2}{2m}\frac{96}{125} (3\pi^2)^{2/3} \\
  &\times\int {\rm d}^3 r \int {\rm d}^3 r' \rho^{5/6}({\bm r})f_1(k_F|{\bm r}-{\bm r}'|) \rho^{5/6}({\bm r}') \\
  & -\frac{\hbar^2}{2m}\int {\rm d}^3 r  \rho^{1/2}({\bm r}) \nabla^2 \rho^{1/2}({\bm r})   \\
  & -\frac{\hbar^2}{2m} \frac{21}{125} (3\pi^2)^{2/3} \int {\rm d}^3 r  \rho^{5/3}({\bm r}),
\end{aligned}
\end{equation}
with
\begin{equation}\label{Eq:f1R}
  f_1(k_F|{\bm r}-{\bm r}'|) = \frac{1}{2\pi^2 R}\int_0^{\infty} l_1(q)\sin(q R)q {\rm d}q.
\end{equation}

Up to here, a non-local (NL) orbital-free KEDF has been constructed, by following the steps of Ref.~\cite{Wang1992PRB}. 
Such a form has never been implemented in calculations of atomic nuclei, and we do it here, in conjunction with the interaction part of a nuclear functional. 
We have adopted functionals of the Skyrme type~\cite{Bender2003Rev.Mod.Phys.}.
Note that, although the derivation is based on a density-independent $k_F$, 
one can still test the non-local, density-dependent $k_F$, 
$k_F=\left[3\pi^2\frac{\rho({\bm r})+\rho({\bm r}')}{2}\right]^{1/3}$,  
by directly incorporating it into Eq.~\eqref{Eq:Tnlfinal}. 
Note that the above derivation of the nonlocal orbital-free KEDF does not involve any specific detail of the interaction. The obtained functional is therefore suitable for various quantum many-body systems.
Indeed, for electronic systems, in quantum chemistry, previous works on nonlocal orbital-free kinetic energy density functionals have already been published. These studies, notably kernel-based formulations~\cite{Wang1992PRB, Smargiassi1994PRB, Wang1999PhysRevB.60.16350, Ho2008PRB, Mi2023CR}, provide valuable references for the development of nonlocal orbital-free density functionals in nuclear physics. Our goal is to implement for the first time nonlocality in the KEDF for nuclear systems to capture quantum shell effects.

\begin{figure}[!htbt]
  \centering
    \includegraphics[width=0.4\textwidth]{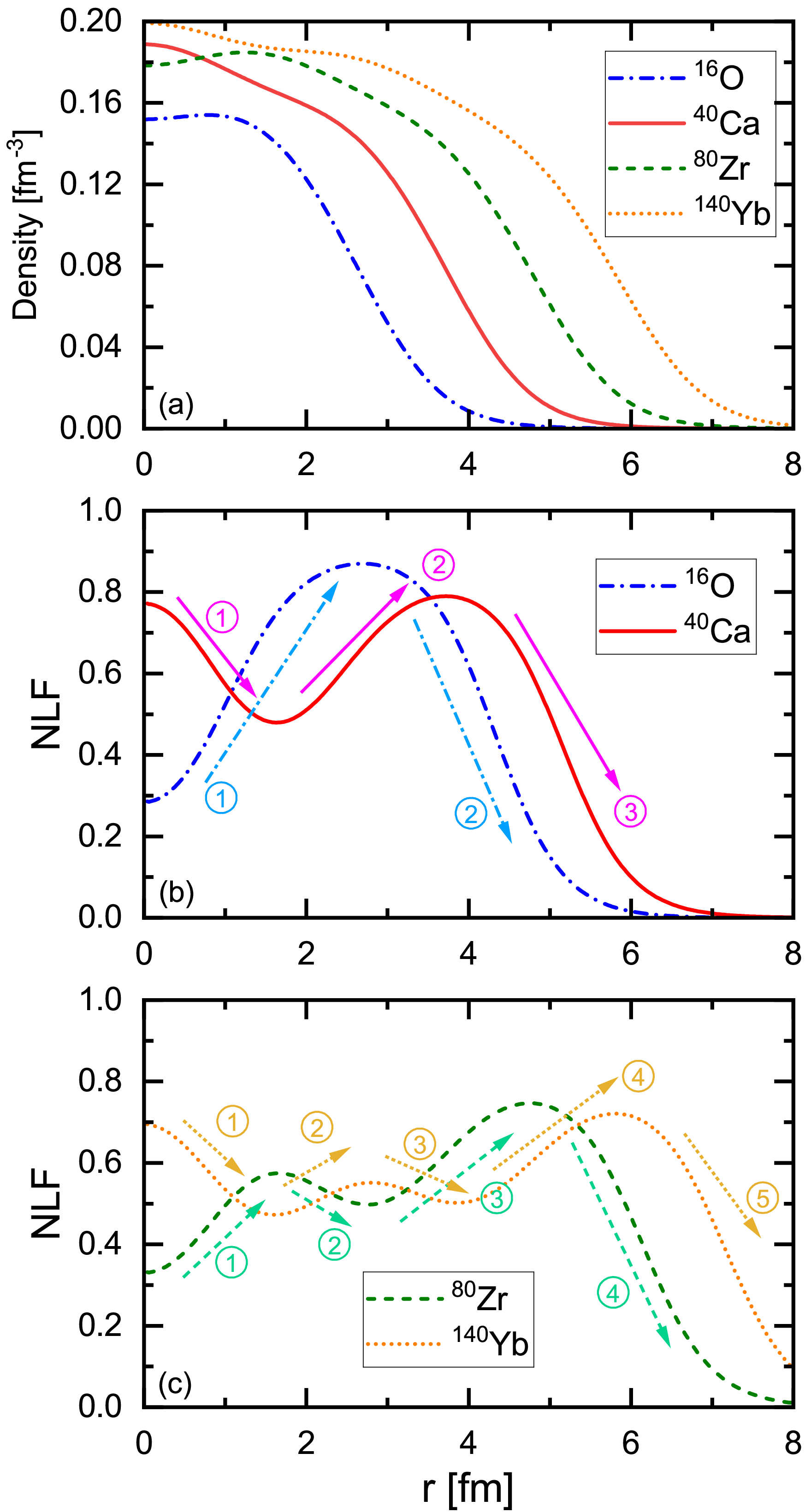}
  \caption{(a) The ground-state densities of $^{16}$O and $^{40}$Ca obtained with the Kohn-Sham 
  approach with the Skyrme SkP interaction. 
  (b) The nucleon localization functions (NLF) of $^{16}$O and $^{40}$Ca obtained with the Kohn-Sham approach.
  (c) Same as (b), but for the $^{80}$Zr and $^{140}$Yb nuclei. 
  In the panels (b) and (c),
  the arrows represent the intervals where the corresponding NLF function is monotonic.
  The circles with numbers are used to count how many times the NLF function changes its monotonic behavior.
  }
\label{fig1}
\end{figure}

{\it Shell effects in the non-local OF-DFT.}
Let us now numerically apply the non-local OF-DFT to the 
$^{16}$O, $^{40}$Ca, $^{80}$Zr, and $^{140}$Yb nuclei, 
which are double magic nuclei in the absence of the spin-orbit interaction (See Supplementary Material~\cite{[See Supplementary Material for more details of the calculations and nuclear major shells] supplement}). 
To this end, we employ the Skyrme SkP interaction \cite{Dobaczewski1984Nucl.Phys.A103}, by ignoring for simplicity not only the spin-orbit interaction but the Coulomb interaction as well.
Notice that the concept of orbitals is absent in the OF-DFT, and thus quantum shell effects should be identified from densities and related quantities.
However, this may not be straightforward.
One may directly look into the fluctuations of the density itself. 
However, as can be seen in Fig.~\ref{fig1}~(a), the existence of shells is far from being easy to spot by looking at the density only.
The same holds for atomic and molecular systems but in that case it has been highlighted that the so-called {\em localization function} works well to extract shell effects from densities~\cite{Becke1990JCP}.
Here, the {\em nucleon localization function}  (NLF) is defined as~\cite{reinhard2011,jerabek2018,zhang2016,Ren2022PRL}
\begin{equation}\label{Eq:NLF}
   {\rm NLF} = \left[1+\left(\frac{\tau-\tau_{\rm vW}}{\tau_{\rm TF}}\right)^2 \right]^{-1},
\end{equation}
where $\tau$ is the kinetic energy density of the system obtained by a target method, 
$\tau_{\rm vW}$ and $\tau_{\rm TF}$ are the kinetic energy densities given by the vW functional and the TF functional, respectively. 

Before we analyze the results of the non-local OF-DFT, we first discuss the result of the 
Kohn-Sham (KS) calculations, which provide testbeds for validating the OF-DFT approaches.
The details of the Kohn-Sham DFT calculations with the shooting method~\cite{Killingbeck1987J.Phys.AMath.Gen.} can be seen in the Supplementary material~\cite{[See Supplementary Material for more details of the calculations and nuclear major shells] supplement}.
As shown in Figs.~\ref{fig1} (b) and (c), the NLF well accounts for the number of shells embedded in the nuclear densities, 
that is, the number of ``up" and ``down" in the NLF curves.
For instance, 
the protons and the neutrons in $^{16}$O occupy the 1$s$-shell and the 1$p$-shell, and thus the number of the occupied major shells is 2 (See Supplementary material~\cite{[See Supplementary Material for more details of the calculations and nuclear major shells] supplement}).
This number coincides with how many times the derivative of the NLF for this nucleus changes its sign. 
This is the case for all the other nuclei shown in Fig. \ref{fig1}, and thus the NLF can be 
employed to benchmark whether the OF-DFT approaches capture shell effects.

\begin{figure}[!htbt]
  \centering
    \includegraphics[width=0.5\textwidth]{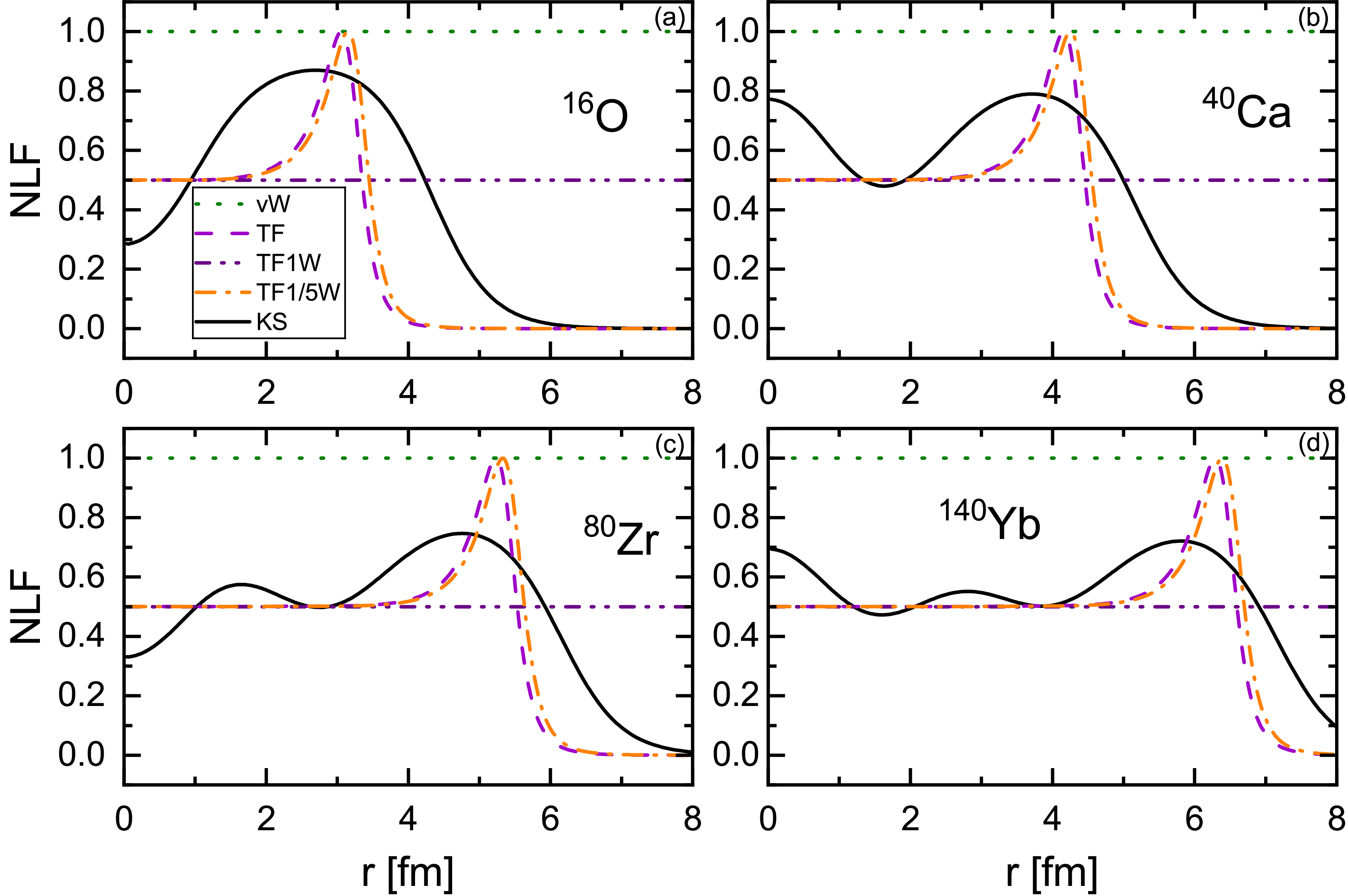}
\caption{The nucleon localization functions for (a) $^{16}$O, (b) $^{40}$Ca, (c) $^{80}$Zr, and (d) $^{140}$Yb 
obtained with the semi-local kinetic energy density functionals, that is, the vW, the TF, the TF1W, 
and the TF1/5W functionals, where $(\alpha,\beta)$ in Eq. (\ref{Eq:TTFvW}) 
is (0,1), (1,0), (1,1) and (1,1/5), respectively. 
For comparison, the figure also shows the nucleon localization function with $\tau(\rho)$ from the Kohn-Sham approach.
}
\label{fig2}
\end{figure}

Figure~\ref{fig2} shows the NLF given by the OF functionals including 
vW, TF, TF1W, and TF1/5W functionals, where $(\alpha,\beta)$ in Eq. (\ref{Eq:TTFvW}) 
is (0,1), (1,0), (1,1) and (1,1/5), respectively, in comparison with the KS results. 
The latter two functionals, together with 
$(\alpha,\beta)=(1,1/9)$, 
are often employed in the orbital-free calculations for atomic systems.
For comparison, the same densities [see Fig.~\ref{fig1}(a)] are used in the evaluation of the NLF from the different orbital-free functionals.
As can be seen in the figure, those semi-local OF-DFT approaches yield 
very different behaviors of the NLF from that with the Kohn-Sham approach for all the four nuclei. 
Note that, according to Eq.~\eqref{Eq:NLF}, the NLFs of the vW and the TF1W functionals are merely constants 
(1 and 0.5, respectively). 
One can observe that the NLFs for the semi-local OF-DFT approaches  
exhibit qualitatively the same behavior across the different nuclei, despite that 
these nuclei should have different shell structures. 
This clearly demonstrates that these semi-local functionals based on the TF and the vW formalisms lack the capacity to capture quantum shell effects.

\begin{figure}[!htbt]
  \centering
    \includegraphics[width=0.5\textwidth]{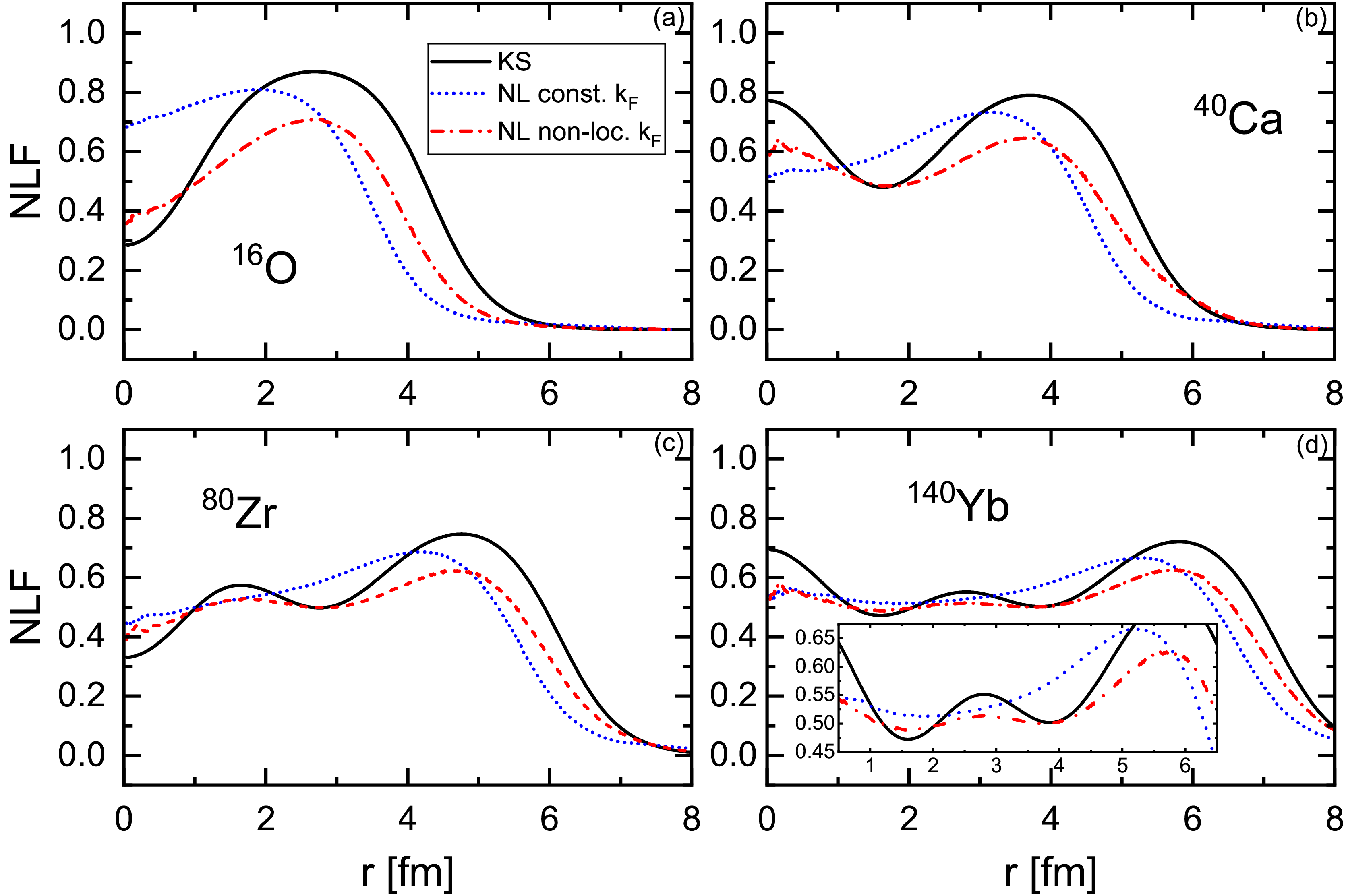}
  \caption{
  Same as Fig. \ref{fig2}, but obtained with the non-local functionals with 
  a constant $k_F$ (the dotted lines) and the density-dependent $k_F$ (the dot-dashed lines). 
  The inset in the panel (d) magnifies a specific part of the figure to better illustrate the detailed behaviors.
  }
\label{fig3}
\end{figure}

The situation completely changes for the non-local (NL) OF functionals. 
Notably, the NL functional with a constant $k_F=0.80~{\rm fm}^{-1}$ (the dotted curves) 
can reproduce the trends of the 
NLFs obtained with the KS approach for all the four nuclei, as shown in Fig.~\ref{fig3}. 
This already represents a notable improvement over other orbital-free functionals, even though
further refinements may still be warranted. 
Note that the adopted $k_F$ has a lower scale than the one corresponding to saturation density, to account for both bulk and surface of nuclear density.
Even more remarkably, as can be seen from Fig.~\ref{fig3}, the NL functional with the density-dependent $k_F$ (the dot-dashed curves) 
can not only reproduce the trend of the KS NLFs, but also reproduce their ``up" and ``down" behaviors, which are associated with quantum shell effects. 
This is a remarkable breakthrough in OF-DFT, as it is the first 
successful orbital-free functional that succeeds in capturing the nuclear quantum shell effects (except 
for the machine-learning approach). 
This would certainly encourage further developments of the non-local orbital-free DFT. 

{\it Summary.} 
We have developed for the first time 
the formalism of a novel non-local kinetic energy density functional for nuclear systems. 
We have focused on a two-point non-local KEDF with a kernel 
determined via the Thomas-Fermi limit as well as the linear-response theory.
This non-local KEDF can be reduced to the TF functional under a condition of a slow density variation, ensuring consistency with the 
established theory. 
To identify shell effects without explicit orbitals in the framework 
of OF-DFT, we have employed the nucleon localization function (NLF). 
We have evaluated the NLFs for $^{16}$O, $^{40}$Ca, $^{80}$Zr, and $^{140}$Yb using the Kohn-Sham densities, 
and found that the NLF effectively captures the number of shells embedded in the nuclear densities: specifically, fluctuations in the NLF (i.e., the number of its ``up and down" variations) correspond to the number of nuclear major shells. 
The TF and the vW functionals, as well as several combinations thereof, produce NLFs that differ significantly from the Kohn-Sham results and fail to reveal distinct nuclear shell structures. In contrast, the non-local OF-DFT with a constant $k_F$ reproduces the trends of the Kohn-Sham NLFs, and notably the non-local OF-DFT with a density-dependent $k_F$ can accurately replicates both the overall trends 
and the NLF fluctuations associated with shell effects,
for all the four nuclei.  

These findings apparently
demonstrate that nuclear quantum shell effects can be explicitly captured in the OF-DFT via a physically interpretable non-local KEDF, 
and thus 
disproves 
a misconception that an OF-DFT is inherently unable to capture quantum shell effects.
This itself is a great advance in the development of the orbital-free methodologies in nuclear physics. 
This can help pave the way for more accurate DFT calculations of e.g., the neutron star inner crust, the superheavy elements, 
and the fission dynamics of heavy nuclei. 

Of course, there are still many ways to improve the calculations presented in this paper.
Firstly, while we have used the density distributions from the Kohn-Sham approach to evaluate the kinetic energy density in the non-local OF-DFT, a more consistent way is to use the self-consistent densities determined variationally with the non-local kernels.
Secondly, one would need to incorporate the spin-orbit and the Coulomb
interactions, which we have ignored for simplicity in this paper, into the non-local OF-DFT approaches (as well as pairing for open-shell systems).
Thirdly, one would need to investigate with the non-local OF-DFT the deformation properties of atomic nuclei, which are closely related to the quantum shell effects.
Lastly, even though we have assumed a particular dependence of the non-local kernel on the density distribution, its actual form has not yet been known. The previous developments and the comparison with experiment may call for other forms of the non-local OF-DFT. 
The machine learning approach may be of help in constructing such forms. 
We leave these to future investigations.



{\it Acknowledgments.}
This work was partly supported by the National Natural Science Foundation of China (Grants Nos. 12405134 and 12475117), National Key R\&D Program of China (Contract No. 2024YFA1612600), JSPS KAKENHI Grant Number JP23K03414, and the High-performance Computing Platform of Peking University.

{\it Data availability.}
The data that support the findings of this article are not publicly available. The data are available from the authors upon reasonable request.

\bibliography{paper}

\end{document}


\begin{CJK*}{UTF8}{} 


\title{Non-local orbital-free density functional theory incorporating nuclear shell effects: Supplementary materials}

\author{Xinhui Wu}
\email{wuxinhui@fzu.edu.cn}
\affiliation{Department of Physics, Fuzhou University, Fuzhou 350108, Fujian, China}

\author{Gianluca Col\`o}
\email{Gianluca.Colo@mi.infn.it}
\affiliation{Dipartimento di Fisica, Universit\`a
degli Studi di Milano, via Celoria 16, 20133 Milano, Italy,}
\affiliation{INFN, Sezione di Milano, via Celoria 16, 20133 Milano, Italy}

\author{Kouichi Hagino}
\email{hagino.kouichi.5m@kyoto-u.ac.jp}
\affiliation{ 
Department of Physics, Kyoto University, Kyoto 606-8502,  Japan} 
\affiliation{Institute for Liberal Arts and Sciences, Kyoto University, Kyoto 606-8501, Japan}
\affiliation{ 
RIKEN Nishina Center for Accelerator-based Science, RIKEN, Wako 351-0198, Japan
}

\author{Pengwei Zhao}
\email{pwzhao@pku.edu.cn}
\affiliation{State Key Laboratory of Nuclear Physics and Technology, School of Physics, Peking University, Beijing 100871, China}

\maketitle

\end{CJK*}

\renewcommand{\thefigure}{S\arabic{figure}}
\renewcommand{\thetable}{S\arabic{table}}

\section{Details of the calculation}

This section provides details of the Kohn-Sham DFT calculations performed in the main text.

In this work, we consider spherical nuclei with the same proton and neutron numbers, and the spin-orbit and the Coulomb interactions are ignored.
The interaction energy $E_{\rm int}$ adopted is the Skyrme functional,
\begin{align}
  E_{\rm int}[\rho(\bm{r})] = &  \int\left( \frac{3}{8}t_{0}\rho^2 + \frac{1}{16}t_{3}\rho^{2+\gamma} \right){\rm d}^3\bm{r} \notag \\
  & + \int \frac{1}{64}(9t_1 - 5t_2 -4t_2x_2)(\nabla\rho)^2 {\rm d}^3\bm{r}, \label{Skyrme}
\end{align}
whose functional derivative reads,
\begin{align}
  \frac{\delta E_{\rm int}[\rho]}{\delta \rho} = & \frac{3}{4}t_{0}\rho + \frac{2+\gamma}{16}t_{3}\rho^{1+\gamma} \notag \\
  & -\frac{1}{32}(9t_{1}-5t_{2}-4t_{2}x_{2})(\Delta\rho). \label{Skyrme_fd}
\end{align}
This latter is the effective mean-field that enters the Kohn-Sham equations.
We take
the Skyrme parameters in Eq.~\eqref{Skyrme}, i.e., $t_0$ $t_1$, $t_2$, $t_3$, $x_1$, and $\gamma$, from the SkP functional~\cite{Dobaczewski1984Nucl.Phys.A103}, which has an effective mass $m^*=m$. 
This ensures that the kinetic term corresponds to that of free nucleons, allowing us to test the kinetic-energy functional in a clean and physically transparent manner, with no need to account for the $\rho\tau$ term in Skyrme functional.
The values of the SkP parameters are
\begin{align}
  & t_0=-2931.70~~\rm{MeV\,fm^3}, \\
  & t_1=320.62~~\rm{MeV\,fm^5}, \\
  & t_2=-337.41~~\rm{MeV\,fm^5}, \\
  & t_3=18708.97~~\rm{MeV\,fm^{3+3\gamma}}, \\
  & x_2=-0.53732, \\
  & \gamma=\frac{1}{6}.
\end{align}
The shooting method is used to solve the radial Schr{\"o}dinger equation in the Kohn-Sham iterations~\cite{Killingbeck1987J.Phys.AMath.Gen., Wu2022Phys.Rev.C}.
The radial coordinate $r$ is discretized on a uniform grid of 501 points from 0 to 20 fm with spacing $\Delta r = 0.04$~fm, i.e.,
\begin{equation}
  r_i = r_0+i\Delta r,~~i=0,1,\cdots,500.
\end{equation}
To avoid the singularity of the centrifugal potential $\dfrac{l(l+1)}{2m_Nr^2}$ at the origin, the value of $r_0$ is taken as small as $\Delta r/10000$.
The boundary conditions for $\psi_{n_rl}(r)$ are taken as
\begin{equation}
  \psi_{n_rl}(r_0) = \psi_{n_rl}(r_{500}) = 0.
\end{equation}

Then, for a given orbital angular momentum $l$, one can either calculate the wavefunctions starting from the ``left'' $\psi^{(l)}_{n_rl}(r_0) = 0$ and $\psi^{(l)}_{n_rl}(r_1) = 1$ to the ``right'' $\psi^{(l)}_{n_rl}(r_i)$ via
\begin{eqnarray}
\psi^{(l)}_{n_rl}(r_{i+1}) &=&
\frac{24-10\Delta r^2f(r_i)}{12+\Delta r^2f(r_{i+1})}  \psi^{(l)}_{n_rl}(r_i) \nonumber \\
&&-\frac{12+\Delta r^2f(r_{i-1})}{12+\Delta r^2f(r_{i+1})} \psi^{(l)}_{n_rl}(r_{i-1}),
\end{eqnarray}
with $f(r)=2m_N\left[\varepsilon_{n_rl} - v(r) - \dfrac{l(l+1)}{2m_Nr^2}\right]$, or from the ``right'' $\psi^{(r)}_{n_rl}(r_{500}) = 0$ and $\psi^{(r)}_{n_rl}(r_{499}) = 1$ to the ``left'' $\psi^{(r)}_{n_rl}(r_i)$, via
\begin{eqnarray}
\psi^{(r)}_{n_rl}(r_{i-1}) &=&
\frac{24-10\Delta r^2f(r_i)}{12+\Delta r^2f(r_{i-1})}  \psi^{(r)}_{n_rl}(r_i) \nonumber \\
&&-\frac{12+\Delta r^2f(r_{i+1})}{12+\Delta r^2f(r_{i-1})} \psi^{(r)}_{n_rl}(r_{i+1}).
\end{eqnarray}
The single-particle energy $\varepsilon_{n_rl}$ can thus be determined by solving the equation at a matching point $m_0$, e.g., $m_0 = 50$,
\begin{equation}
  {\rm Det}(\varepsilon_{n_rl})\equiv
  \left|
    \begin{matrix}
      \psi^{(r)}_{n_rl}(r_{m_0-1})  &\psi^{(r)}_{n_rl}(r_{m_0})\\
      \psi^{(l)}_{n_rl}(r_{m_0-1})  &\psi^{(l)}_{n_rl}(r_{m_0})\\
    \end{matrix}
  \right|=0.
\end{equation}
In practice, this equation is solved by the bisection search for the single-particle energy $\varepsilon_{n_rl}$.
The obtained eigenfunction is then normalized and further used to calculate various quantities.

\section{Nuclear major shells}

\begin{figure*}[!htbt]
  \centering
  \includegraphics[width=0.8\textwidth]{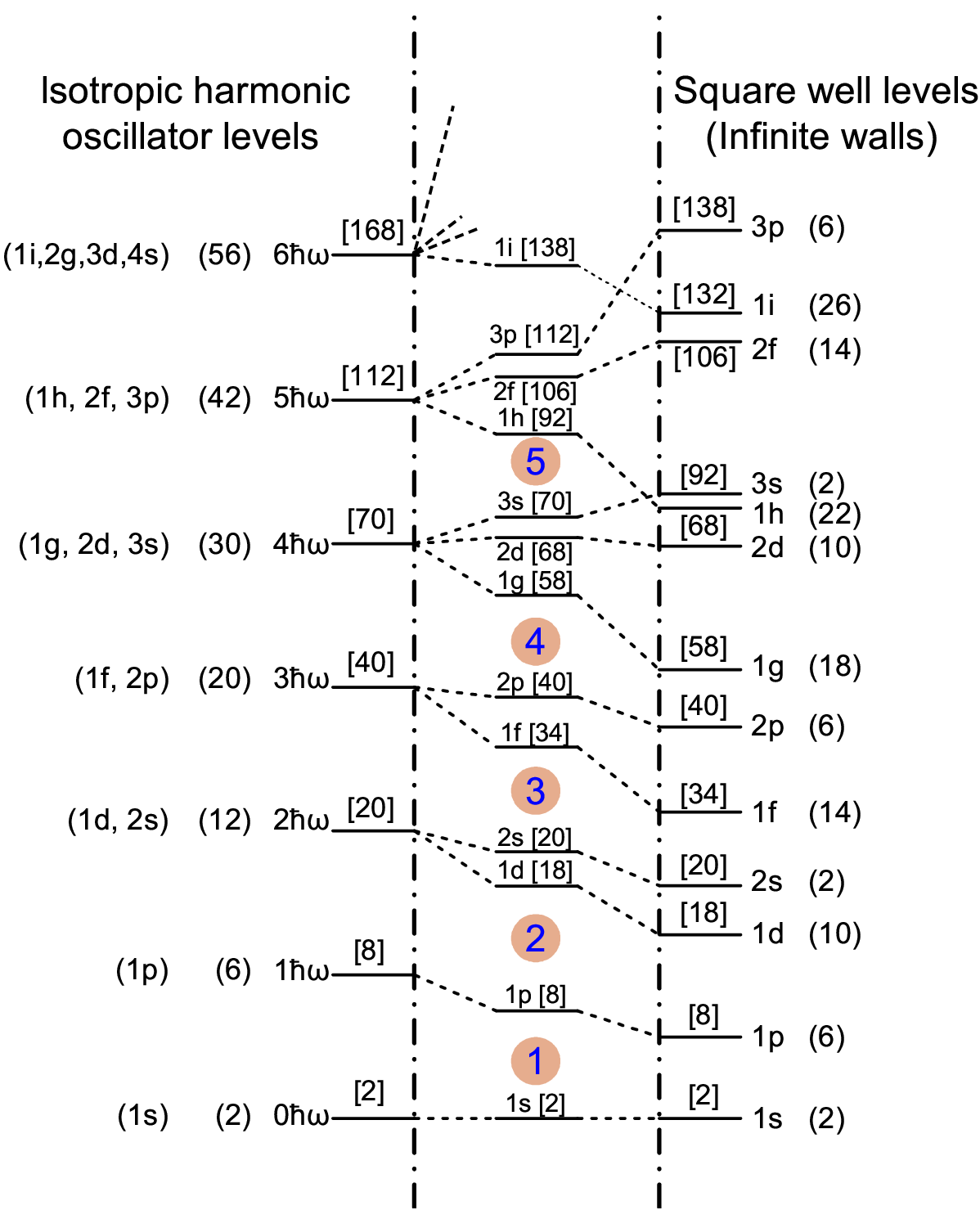}
  \caption{Level scheme of an isotropic harmonic oscillator (l.h.s) and of an infinite square well (r.h.s). The actual level scheme from the Kohn-Sham calculation is shown in the middle. The numbered circles label the occupied major shells in the nuclei discussed in the main text.
    }
\label{supfig1}
\end{figure*}

This short section provides some simple materials, yet useful for the non-expert reader, that highlights the nuclear shell structure from the underlying single-particle level scheme.
Fig.~\ref{supfig1} presents the single-particle level scheme of an isotropic harmonic oscillator (left) and of an infinite square well (right), as well as the one from the Kohn-Sham calculation of $^{140}$Yb with the SkP functional (middle).
Note that spin-orbit and the Coulomb interactions are ignored in the present study.
The levels in the realistic level scheme from the Kohn-Sham calculation mostly lie between the ones of an isotropic harmonic oscillator and the ones of an infinite square well.

One can see that, in the level scheme from the Kohn-Sham calculation, several distinct nuclear major shells emerge, corresponding to the energy gaps between groups of single-particle states.
Specifically, five major shells can be clearly identified (i.e., 2, 8, 20, 40, 70), which are labeled by the numbers 1-5 in Fig.~\ref{supfig1}.
These numbers correspond one-to-one to the shell indices marked in Fig.~1 of the main text, where the same numbers were used to indicate the changes in the monotonic behavior of the nucleon localization function (NLF).
This correspondence confirms that the oscillatory pattern of the NLF effectively reflects the underlying shell closures in the self-consistent single-particle spectrum.

\bibliography{paper}